\documentclass[twocolumn,aps,prb]{revtex4}
\usepackage{graphicx}
\usepackage{amssymb}
\usepackage{amsmath}
\usepackage{bm}
\usepackage{color}

\def\Journal #1,#2,#3,#4#5#6#7{#1 {\bf #2}, #3 (#4#5#6#7)}

\def\e{\varepsilon}
\def\p{\prime}
\def\s{\sigma}

\begin{document}

\title{Quantum transport in three-dimensional Weyl electron system 
	--- in the presence of charged impurity scattering}
\author{Yuya Ominato and Mikito Koshino}
\affiliation{Department of Physics, Tohoku University, Sendai 980-8578, Japan}
\date{\today}

\begin{abstract}
We theoretically study the quantum transport in three-dimensional Weyl electron system
in the presence of the charged impurity scattering using a self-consistent Born approximation (SCBA).
The scattering strength is characterized by
the effective fine structure constant $\alpha$, 
which depends on the dielectric constant and the Fermi velocity of the linear band.
We find that the Boltzmann theory fails at the band touching point,
where the conductivity takes a nearly constant value
almost independent of $\alpha$, even though 
the density of states linearly increases with $\alpha$.
There the magnitude of the conductivity only depends on the impurity 
density.
The qualitative behavior is quite different from the case of the Gaussian impurities,
where the minimum conductivity vanishes below a certain critical 
impurity strength.
\end{abstract}

\maketitle

\section{Introduction}

The electronic property of the three-dimensional (3D)
gapless system is one of the great interest
in the recent condensed matter physics.
There two diffrent energy bands 
stick together at isolated points in the Brillouin zone, 
and the electronic structure around each touching point 
is described by the Weyl Hamiltonian.
There are several theoretical proposals for 
possible physical systems having gapless band structure,
\cite{burkov2011weyl,burkov2011topological,wan2011topological,young2012dirac,wang2012dirac,singh2012topological,smith2011dirac,liu2013chiral,witczak2012topological,xu2011chern,cho2012possible,halasz2012time}
and in recent experiments, the gapless band structure
was observed in $\rm{Cd}_3\rm{As}_2$ and $\rm{Na}_3\rm{Bi}$
by a angle-resolved photoemission spectroscopy.
\cite{PhysRevLett.113.027603,neupane2014observation,liu2014discovery}

In this paper, we study the electronic transport
in the 3D Weyl electron system
with the charged (Coulomb) impurities.
The impurity effects and the transport properties 
in the 3D gapless electronic system
have been studied in several theoretical works.
\cite{fradkin1986critical1,fradkin1986critical2,burkov2011weyl,burkov2011topological,hosur2012charge,PhysRevLett.112.016402,nandkishore2014rare,biswas2014diffusive,ominato2014quantum,PhysRevLett.113.026602,skinner2014coulomb,hwang2014carrier,syzranov2014critical}
Previously we studied the conductivity
of single-flavored 3D Weyl system assuming Gaussian impurities,
and found that there is a certain critical disorder strength
at which the conductivity significantly changes its behavior.
\cite{PhysRevLett.112.016402,nandkishore2014rare,biswas2014diffusive,ominato2014quantum,PhysRevLett.113.026602}
The specific form of the impurity potential,
however, generally affects the qualitative behavior
of the electronic transport, and one may ask how the characteristic features
in Gaussian impurities are modified in other types of the scatterers,
such as the typical Coulomb impurities.
For  graphene, i.e., the two-dimensional version of the Weyl electron,
the conductivity was calculated in different scattering models such as
short-ranged impurities \cite{shon1998quantum},
Coulomb impurities \cite{ando2006screening,nomura2006quantum}, and 
the Gaussian impurities, \cite{noro2010theory}
and there the qualitative difference was found 
in the Fermi energy dependence and also in the behavior at the band touching point.
For 3D Weyl system, the effect of Coulomb impurity on the transport 
was studied in the Boltzmann approach. \cite{skinner2014coulomb,burkov2011topological,hwang2014carrier}
Quite recently, the conductivity at the band touching point
in presence of Coulomb impurities was calculated
using the Boltzmann approach together with
the electron-hole puddle picture. \cite{skinner2014coulomb}

When we consider the conductivity near the Weyl point (band touching
point), it is nontrivial
how to appropriately incorporate the finite level broadening effect.
Here we calculate the conductivity of the 3D Weyl electron system
in the presence of the charged Coulomb impurities,
by using the self-consistent Born approximation to treat the finite level
broadening,
and including the screening effect within the Thomas Fermi approximation. 
The scattering strength is characterized by
the effective fine structure constant $\alpha$, 
which depends on the dielectric constant and the Fermi velocity of the linear band.
We find that the density of states is enhanced in all energy region
linearly with the increase of $\alpha$.
On the other hand the conductivity at the Weyl point is almost independent of
$\alpha$ unlike the Boltzmann theory, and even survive in the weak scattering limit, $\alpha\to0$.
The conductivity approaches the Boltzmann theory
away from the Weyl point, 
as long as the Fermi energy is greater than the broadening energy.
The qualitative behavior is quite different 
from the Gaussian impurities,
where the Weyl-point conductivity jumps from zero to a finite value
at some critical scattering strength.
We closely argue about the criteria for 
the critical behavior in general impurity potential under the screening effect.

The paper is organized as follows.
In Sec.\ \ref{sec_form}, we introduce the model Hamiltonian,
and present the formalism to calculate the Boltzmann conductivity
and the SCBA conductivity.
In Sec.\ \ref{sec_app},
we derive an approximate solution of SCBA equation 
at zero energy,
and In Sec.\ \ref{sec_num},
we present the numerical results of the SCBA equation,
for the conductivity and the density of states.
In Sec.\ \ref{sec_dis}, we discuss about 
the validity of the SCBA, which is particularly nontrivial at the Weyl point.
We also argue about the qualitative difference
between different types of the impurity potential.
A brief summary is given in Sec.\ \ref{sec_conc}.

\section{Formulation}
\label{sec_form}

\subsection{Hamiltonian}
We consider a three-dimensional, single-node Weyl
electron system described by a Hamiltonian,
\begin{align}
\mathcal{H}=\hbar v\bm{\s}\cdot\bm{k} + \sum_j U(\bm{r}-\bm{r}_j),
\end{align}
where $\bm{\s}=(\s_x,\s_y,\s_z)$ is the Pauli matrices, $\bm{k}$ is
a wave vector, and $v$ is a constant Fermi velocity.
The first term is the Weyl Hamiltonian, and the second term is the disorder potential
where $\bm{r}_j$ is the positions of randomly distributed scatterers.
For each single scatterers, we assume a long-ranged screened Coulomb potential,
\begin{align}
&U(\bm{r})=\pm\frac{e^2}{\kappa r}
\exp\left(-q_{\rm s}r\right),
\label{eq_imp_potential}
\end{align}
where $\kappa$ is the static dielectric constant,
and scatterers of $\pm$ are randomly distributed with equal probability,
and $q_{\rm s}$ is the Thomas-Fermi screening constant given by
\begin{align}
q_{\rm s}^2=\frac{4\pi e^2}{\kappa}D(\e_{\rm F}),
\label{eq_TF}
\end{align}
at zero temperature.
$U$ is Fourier transformed as
$U(\bm{r})=\int{{\rm d}\bm{q}}u(\bm{q})e^{i\bm{q}\cdot\bm{r}}/{(2\pi)^3}$ where
\begin{align}
u(\bm{q})=\pm \frac{4\pi e^2}{\kappa(q^2+q_{\rm s}^2)}.
\end{align}
We introduce an effective fine-structure constant
\begin{align}
\alpha=\frac{e^2}{\kappa\hbar v},
\end{align}
which characterizes the scattering strength.
For the 3D Weyl electron in $\rm{Cd}_3\rm{As}_2$,
for example, $\alpha$ is estimated at about 0.06
from $v\approx1.0\times10^6{\rm ms}^{-1}$ and $\kappa\approx36$.
\cite{neupane2014observation,PhysRevLett.113.027603,jeon2014landau,jay1977electron}
We define a wave vector scale and an energy scale,
\begin{align}
q_0&=n_{\rm i}^{1/3}, \\
\e_0&=\hbar v q_0,
\end{align}
where $n_{\rm i}$ is the number of scatterers per unit volume.

\subsection{Boltzmann transport theory}

\begin{figure}
\begin{center}
\leavevmode\includegraphics[width=0.8\hsize]{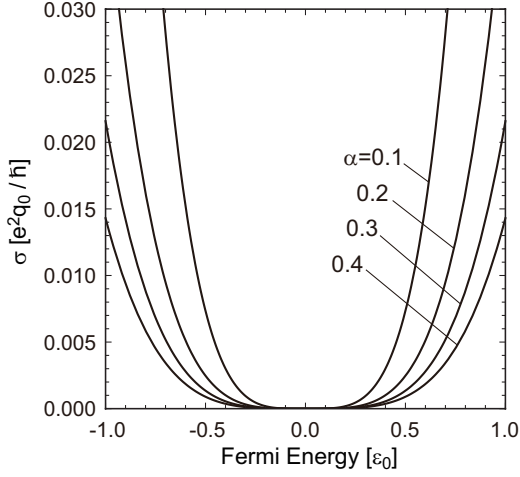}
\end{center}
\caption{Boltzmann conductivity [Eq.\ (\ref{Boltzmann})]
plotted as a function of the Fermi energy.
}
\label{boltz_fig}
\end{figure}

The Boltzmann transport equation for the distribution function
$f_{s\bm{k}}$ is given by
\begin{align}
-e\bm{E}\cdot\bm{v}_{s\bm{k}}\frac{\partial f_{s\bm{k}}}{\partial\e_{s\bm{k}}}
        =\sum_{s^\p}\int\frac{{\rm d}\bm{k}^\p}{(2\pi)^3}(f_{s^\p\bm{k}^\p}-f_{s\bm{k}})W_{s^\p\bm{k}^\p,s\bm{k}},
         \label{Boltzmanneq}
\end{align}
where $s=\pm1$ is a label for conduction and valence bands, and
$W_{s^\p\bm{k}^\p,s\bm{k}}$ is the scattering probability,
\begin{align}
W_{s^\p\bm{k}^\p,s\bm{k}}=
&\frac{2\pi}{\hbar}n_{\rm i}
|\langle s' \bm{k}'| U | s \bm{k} \rangle |^2
\delta(\e_{s^\p\bm{k}^\p}-\e_{s\bm{k}}).
\end{align}
The conductivity is obtained by solving Eq.\ (\ref{Boltzmanneq}).
As usual manner, the transport relaxation time 
 $\tau_{\rm tr}$ is defined by
\begin{align}
\frac{1}{\tau_{\rm tr}(\e_{s\bm{k}})}=\int\frac{{\rm d}\bm{k}^\p}{(2\pi)^3}(1-\cos\theta_{\bm{k}\bm{k}^\p})W_{s\bm{k}^\p,s\bm{k}},
\end{align}
where $\theta_{\bm{k}\bm{k}^\p}$ is the angle between $\bm{k}$ and $\bm{k}^\p$.
For the isotropic scatterers, i.e., $u(\bm{q})$ 
depending only on $q=|\bm{{q}}|$, 
it is straightforward to show that $\tau_{\rm tr}(\e_{s\bm{k}})$ solely depends
on the energy $\e$ and written as \cite{burkov2011topological,ominato2014quantum}
\begin{align}
\frac{1}{\tau_{\rm tr}(\e)}=&
\frac{\pi}{\hbar}n_{\rm i} D_0(\e)
\int_{-1}^{1} {\rm d}(\cos\theta) \, 
u^2[2k\sin(\theta/2)]
\notag\\
&\qquad\qquad \times
(1-\cos\theta)\frac{1+\cos\theta}{2},
\end{align}
where $k = \e/(\hbar v)$ and
$D_0(\e)$ is the density of states in the ideal Weyl electron,
\begin{align}
D_0(\e)=\frac{\e^2}{2\pi^2(\hbar v)^3}.
\end{align}
For the Coulomb impurities,
the relaxation time is derived analytically and written as
\begin{align}
\tau_{\rm tr}(\e)=\frac{ \e^2 }
                               { 4\pi\hbar^2 v^3n_{\rm i} }
                        h(\alpha),
\end{align}
where
\begin{align}
h(\alpha)=\frac{1}{\alpha^2}\left[
                          \left(
                          1+\frac{\alpha}{\pi}
                          \right)
                          \tanh^{-1}\left(
                                        \frac{1}{1+\alpha/\pi}
                                        \right)
                         -1
                         \right]^{-1}.
\end{align}
The conductivity at $T=0$ is given by
\begin{align}
\s_{\rm B}(\e)=e^2\frac{v^2}{3}D_0(\e)\tau_{\rm tr}(\e),
\end{align}
and written as
\begin{align}
\s_{\rm B}(\e)=\frac{1}{24\pi^3}\frac{e^2 q_0}{\hbar}
\left(\frac{\e}{\e_0}\right)^4h(\alpha).
\label{Boltzmann}
\end{align}
Since the electron concentration $n$ is proportional to $\e_{\rm F}^3$
in the 3D linear band,
the Boltzmann conductivity $\s_{\rm B}$ is proportional to $n^{4/3}$.
Figure \ref{boltz_fig} shows the conductivity
Eq.\ (\ref{Boltzmann}) versus the Fermi energy $\e_{\rm F}$ for several values of $\alpha$.

The Boltzmann conductivity in 3D Weyl electron was previously calculated 
under the conditon that the electron density is equal to the Coulomb impurity density,
i.e., all carriers are supplied from the ionic impurities. \cite{burkov2011topological}
The result is reproduced by Eq.\ (\ref{Boltzmann}) with $\e$ is replaced with
$\hbar v (6\pi^2 n_{\rm i})^{1/3}$.


\subsection{Self-consistent Born approximation}

We introduce the self-consistent Born approximation (SCBA)
for 3D Weyl electron system, following 
the formulation for general isotropic impurity potential. \cite{ominato2014quantum}
We define the averaged Green's function as
\begin{align}
\hat G(\bm{k},\e) = \biggl\langle\frac{1}{\e-\mathcal{H}}\biggr\rangle 
       = \frac{1}{\e - \hbar v\bm{\sigma}\cdot\bm{k} - \hat \Sigma(\bm{k},\e)},
\label{def_Gf}
\end{align}
where $\langle\cdots\rangle$ represents the average over
the configuration of the impurity position.
$\hat\Sigma(\bm{k},\e)$ is the self-energy matrix,
which is approximated in SCBA as
\begin{align}
\hat \Sigma(\bm{k},\e)=\int\frac{{\rm d}\bm{k}^\p}{(2\pi)^3}n_{\rm
 i}|u(\bm{k}-\bm{k}^\p)|^2\hat G(\bm{k}^\p,\e).
\label{eq_self}
\end{align}
Eqs.\ (\ref{def_Gf}) and (\ref{eq_self})
are a set of equations to be solved self-consistently.
From the symmetry of the present system, the self-energy matrix can
be expressed as

\begin{align}
\hat \Sigma(\bm{k},\e)=\Sigma_1(k,\e)+\Sigma_2(k,\e)(\bm{\sigma}\cdot\bm{n}),
\label{self_energy}
\end{align}
where $k=|\bm{k}|$ and $\bm{n}=\bm{k}/k$.
We define $X(k,\e)$ and $Y(k,\e)$ as
\begin{align}
X(k,\e)&=\e-\Sigma_1(k,\e), \\ 
Y(k,\e)&=\hbar v k +\Sigma_2(k,\e).
\label{def_XY}
\end{align}

Substituting Eq.\ (\ref{def_Gf}) for Eq.\ (\ref{eq_self}),
the self-consistent equation becomes,
\begin{align}
X(k,\e)&=\e-\int_0^\infty\frac{{k^\p}^2{\rm d}k^\p}{(2\pi)^3}n_{\rm i}V_0^2(k,k^\p)
                                      \frac{X^\p}{{{X^\p}^2-{Y^\p}^2}},
\label{eq_self_x} \\
Y(k,\e)&=\hbar v k +\int_0^\infty\frac{{k^\p}^2{\rm d}k^\p}{(2\pi)^3}n_{\rm i}V_1^2(k,k^\p)
                                      \frac{Y^\p}{{{X^\p}^2-{Y^\p}^2}}.
\label{eq_self_y}
\end{align}
where $X^\p=X(k^\p,\e)$, $Y^\p=Y(k^\p,\e)$, and
\begin{align}
V_n^2(k,k^\p)=2\pi\int^1_{-1}{\rm d}(\cos\theta_{\bm{k}\bm{k}^\p})
                     |u(\bm{k}-\bm{k}^\p)|^2\cos^n\theta_{\bm{k}\bm{k}^\p}.
\label{vn}
\end{align}
The detail of the derivation of 
Eq.\ (\ref{eq_self_x}) and Eq.\ (\ref{eq_self_y}) is given in the Appendix.
From the obtained Green's function,
the density of states per unit area is calculated as
\begin{align}
D(\e)=-\frac{1}{\pi}{\rm Im}\int\frac{{\rm d}\bm{k}}{(2\pi)^3}{\rm Tr}[\hat G(\bm{k},\e+i0)].
\label{eq_dos}
\end{align}

The current vertex part $J_n$ satisfy the Bethe-Salpeter equation
\begin{align}
   \begin{pmatrix}
    J_0  \\
    J_1  \\
    J_2  \\
    J_3   
   \end{pmatrix}=\begin{pmatrix}
                        1  \\
                        0  \\
                        0  \\
                        0   
                       \end{pmatrix}+&\int_0^\infty\frac{{k^\p}^2{\rm d}k^\p}{(2\pi)^3}
                                           \frac{n_{\rm i}}{(X^2-Y^2)({X^\p}^2-{Y^\p}^2)} \notag \\
            &\times\begin{pmatrix}
                        V_0^2 & -(V_0^2-V_2^2)/2 & 0 & 0 \\
                        0 & -(V_0^2-3V_2^2)/2  & 0 & 0 \\
                        0 & 0 & V_1^2 & 0 \\
                        0 & 0 & 0 & V_1^2
                       \end{pmatrix} \notag \\
            &\times\begin{pmatrix}
                        XX^\p & YY^\p & YX^\p & XY^\p \\
                        YY^\p & XX^\p & XY^\p & YX^\p \\
                        YX^\p & XY^\p & XX^\p & YY^\p \\
                        XY^\p & YX^\p & YY^\p & XX^\p
                       \end{pmatrix}    \begin{pmatrix}
                                              J_0^\p  \\
                                              J_1^\p  \\
                                              J_2^\p  \\
                                              J_3^\p       
                                             \end{pmatrix},
\label{eq_self_J}
\end{align}
where $X=X(k^\p,\e)$, $X^\p=X(k^\p,\e^\p)$, $J_0=J_0(k,\e,\e^\p)$, $J_0^\p=J_0(k^\p,\e,\e^\p)$, etc.
The conductivity is calculated with the following formula
\begin{align}
\s(\e)=&\frac{4\hbar e^2v^2}{3}\int_0^\infty\frac{k^2{\rm d}k}{(2\pi)^3} \notag \\
                     \times{\rm Re}&\biggl[\frac{1}{|X^2-Y^2|^2} \notag \\
                             &\times\Bigl\{(3|X|^2-|Y|^2)J_0^{+-}+(3|Y|^2-|X|^2)J_1^{+-} \notag \\
                             &+(3YX^\ast-XY^\ast)J_2^{+-}+(3XY^\ast-YX^\ast)J_3^{+-}\Bigr\} \notag \\
                             &-\frac{1}{(X^2-Y^2)^2} \notag \\
                             &\times\Bigl\{(3X^2-Y^2)J_0^{++}+(3Y^2-X^2)J_1^{++} \notag \\
                             &+2XYJ_2^{++}+2XYJ_3^{++}\Bigr\}\biggr],
\label{eq_cond}
\end{align}
where $X=X(k,\e+i0)$, $J_0^{ss'} = J_0(k,\e+is0,\e+is'0)$, etc. 
The derivation of Eq.\ (\ref{eq_self_J}) and Eq.\ (\ref{eq_cond})
is presented in the Appendix.

\section{Approximate analytical solution at zero energy}
\label{sec_app}

In this section, we derive approximate analytical expressions
for the density of states and the conductivity
at the Weyl point $(\e=0)$.
In the following, we solve the self-consistent Eqs.\ (\ref{eq_self_x})
and (\ref{eq_self_y}) at $\e=0$
using a certain approximation to simplify the problem.
We first assume $Y(k,0)$ is written as
\begin{align}
Y(k,0)=\hbar v k,
\label{y_approx}
\end{align}
i.e., we neglect the $\Sigma_2$ term in Eq.\ (\ref{def_XY}).
We can show that $\Sigma_2$ is also linear to $k$ in the real solution,
and thus it gives Fermi velocity renormalization,
while it does not change the qualitative behavior
of the density of states and the conductivity.
Then the equation \ (\ref{eq_self_x}) for $X(k)=X(k,0)$
is written as
\begin{align}
X(k)&=\-\int_0^\infty\frac{{k^\p}^2{\rm d}k^\p}{(2\pi)^3}n_{\rm i}V_0^2(k,k^\p)
\frac{X(k^\p)}{{(X(k^\p))^2-(\hbar v k^\p)^2}},
\label{eq_self_x2}
\end{align}
where
\begin{align}
 V^2_0(k,k^\p)&=\left(\frac{4\pi e^2}{\kappa}\right)^2
\frac{ 4\pi }
{(k^2 + {k^\p}^2 + q_{\rm s}^2)^2-4k^2{k^\p}^2}.
\end{align}

First we consider the solution $X(k)$ in $k \gg q_{\rm s}$.
$V_0^2(k,k')$ can be approximately written by a delta function as
\begin{align}
{k^\p}^2 V^2_0(k,k^\p)\approx\left(\frac{4\pi e^2}{\kappa}\right)^2
\frac{\pi^2}{q_{\rm s}}\delta(k-k^\p),
\label{mat_approx}
\end{align}
and Eq.\ (\ref{eq_self_x2}) then becomes
\begin{align}
X(k)=-\frac{n_{\rm i}}{(2\pi)^3}
\left(\frac{4\pi e^2}{\kappa}\right)^2
\frac{\pi^2}{q_{\rm s}}
\frac{X(k)}{(X(k))^2-(\hbar v k)^2}.
\end{align}
The physically plausible solution is
\begin{align}
X(k)= 
\left\{
\begin{array}{cc}
i\sqrt{\Gamma_0^2-(\hbar vk)^2} & (k<\Gamma_0/(\hbar v)), \\
0 & (k > \Gamma_0/(\hbar v)), 
\end{array}
\right.
\label{x_approx2}
\end{align}
where
\begin{align}
	\Gamma_0 = \hbar v q_0 \sqrt{\dfrac{2\pi\alpha^2}{(q_{\rm s}/q_0)}}.
\label{gamma0}
\end{align}
Therefore, $X(k)$ attenuates with the increase of $k$
and vanishes at $k=\Gamma_0/(\hbar v)$.

For $k=0$, we need a special treatment since the approximation Eq.\ (\ref{mat_approx})
is not valid in $k < q_{\rm s}$.
The self-consistent equation at $k=0$ is written as
\begin{align}
X(0)=-\int^\infty_0\frac{{k^\p}^2{\rm d}k^\p}{(2\pi)^3}n_{\rm i}
&\left(\frac{4\pi e^2}{\kappa}\right)^2
\frac{4\pi}{({k^\p}^2+q_{\rm s}^2)^2} \notag \\
&\times\frac{X(k^\p)}{{(X(k^\p))^2-(\hbar v k^\p)^2}}.
\end{align}
On the condition that $\Gamma_0 \gg \hbar v q_{\rm s}$,
the term $({k^\p}^2+q_{\rm s}^2)^{-2}$ is a rapidly changing function
compared to $X(k')$, and it vanishes except in the vicinity of $k'=0$.
Then $X(k')$ can be replaced by $X(0)$  in the integral,
and we obtain a solution,
\begin{align}
 X(0) = i \Gamma ,
\end{align}
with
\begin{align}
 \Gamma= \Gamma_0 - \hbar v q_{\rm s}.
\label{gamma}
\end{align}
When compared to Eq.\ (\ref{x_approx2}),
we notice that $X(0)$ has an additional correction term $-i\hbar v q_{\rm s}$,
which is actually important in considering the limit of $\alpha\to 0$.
All the approximation above is based on the assumption $\Gamma_0 \gg \hbar v q_{\rm s}$,
and this is actually satisfied in the situation considered 
in the later sections. 


Based on the above arguments, we introduce a crude approximation
by even simplifying $X(k)$ to a step function as
\begin{align}
X(k)=\begin{cases}
        i\Gamma \hspace{4mm} (k<\Gamma/(\hbar v))  \\
        0 \hspace{5.6mm} (k>\Gamma/(\hbar v))
       \end{cases},
\label{x_approx3}
\end{align}
with $\Gamma$ defined in Eq.\ (\ref{gamma}). 
Substituting Eq.\ (\ref{y_approx}) and (\ref{x_approx3})
for Eq.\ (\ref{eq_dos}), we find the density of states 
\begin{align}
&D(\e=0)=\frac{\Gamma^2}{(\hbar v)^3}\frac{f}{4\pi}, \label{dos_at_zero} \\
&f=\frac{4-\pi}{\pi^2}\approx 0.087,
\end{align}
and from Eq.\ (\ref{eq_TF}), the screening constant is written as
\begin{align}
q_{\rm s}=\frac{\Gamma}{\hbar v}\sqrt{f\alpha}.
\label{eq_TF2}
\end{align}
By solving Eq.\ (\ref{gamma}) and (\ref{eq_TF2}), we have
\begin{align}
\Gamma=\e_0\left(
                           \frac{2\pi}
                                  {\sqrt{f}(1+\sqrt{f\alpha})^2} 
                      \right)^{1/3}\sqrt{\alpha}.
\label{q_g_approx}
\end{align}

In $\alpha\ll1$, 
$\Gamma$ is nearly proportional to $\sqrt{\alpha}$
and 
the density of states is proportional to $\Gamma^2$, thus to $\alpha$.

%

The Bethe-Salpeter equation Eq.\ (\ref{eq_self_J}) 
can be approximately solved at $\e=0$ in a similar manner.
We assume the form of the solution as,
\begin{align}
   \begin{pmatrix}
    J_0^{+s}(k)  \\
    J_1^{+s}(k)  \\
    J_2^{+s}(k)  \\
    J_3^{+s}(k)   
   \end{pmatrix}& \approx \begin{pmatrix}
                                    J_0^{+s}(k)  \\
                                    0  \\
                                    0  \\
                                    0   
                                   \end{pmatrix},
\label{vertex_approx}
\end{align}
where $s=\pm$.
Then the equation is reduced to
\begin{align}
J_0^{+s}(k)=1+&\int_0^\infty\frac{{k^\p}^2{\rm d}k^\p}{(2\pi)^3}
                 \frac{n_{\rm i}}{(X^2-Y^2)({X^\p}^2-{Y^\p}^2)} \notag \\
               &\times \left(V_0^2XX^\p-\dfrac{V_0^2-V_2^2}{2}YY^\p\right)J_0^{+s}(k').
\end{align}
In a similar manner to $X(k)$,
we find a solution,
\begin{align}
J_0^{+s}(k)=\begin{cases}
             J^{+s} \hspace{4mm} (k<\Gamma/(\hbar v))  \\
             0 \hspace{8mm} (k>\Gamma/(\hbar v))
            \end{cases},
\end{align}
where
\begin{align}
J^{+s}&=    
                \left[
                1+s\frac{2\pi\alpha^2(3q_{\Gamma}-{q}_{\rm s})q_0^3}
                              {3q_{\rm s}(q_{\Gamma}+q_{\rm s})^3}
                \right]^{-1},
\label{J_approx}
\end{align}
and $q_{\Gamma}=\Gamma/(\hbar v)$.
In $\alpha\ll1$, $J^{+s}$ can be expanded in the lowest order of $\alpha$ as
\begin{align}
&J^{+-}\approx\frac{3}{4\sqrt{f}}
\frac{1}{\sqrt{\alpha}},
\quad J^{++}\approx\frac{1}{2},
\label{eq_J_expand}
\end{align}
i.e., $J^{+-}$ diverges in $\alpha\to 0$ while $J^{++}$ remains constant.
In small $\alpha$, therefore, we can neglect $J^{++}_0$ in Eq.\ (\ref{eq_cond})
leaving only $J^{+-}_0$, and then the conductivity is calculated as
\begin{align}
\s(\e=0)
        \approx&\frac{4\hbar e^2v^2}{3}\int_0^{\Gamma/(\hbar v)}\frac{k^2{\rm d}k}{(2\pi)^3}
                     \frac{3J^{+-}}{\Gamma^2}
                     \notag\\
=&\frac{J^{+-}}{6\pi^3}\frac{e^2}{\hbar}\frac{\Gamma}{\hbar v}.
\label{cond_approx}
\end{align}
Using Eq.\ (\ref{eq_J_expand}), the conductivity in the limit of $\alpha\to 0$
becomes
\begin{align}
\sigma(\e=0)&\approx\frac{1}{8\pi^3}
                    \left(
                    \frac{2\pi}{f^2}
                    \right)^{1/3}
                    \frac{e^2 q_0}{\hbar} \notag \\
          &\approx0.038\times \frac{e^2 q_0}{\hbar}.
          \label{cond_approx2}
\end{align}
Here the magnitude of the conductivity is determined solely by
the impurity density $n_{\rm i} = q_0^3$,
and it scales in proportion to $n_{\rm i}^{1/3}$.

The conductivity formula Eq.\ (\ref{cond_approx}) is almost  
equivalent to the analytical expression for the Gaussian impurities \cite{ominato2014quantum}, 
but the actual behavior of the conductivity is significantly different.
In the Gaussian case, the vertex part $J^{+-}$ is constant and
the level broadening $\Gamma$ vanishes below the critical disorder strength.
As a result, the conductivity vanishes in the weak disorder regime.
In the Coulomb impurity case, on the other hand,
$J^{+-}$ diverges as $1/\sqrt{\alpha}$
in the limit of $\alpha\to0$, while the level broadening vanishes as $\sqrt{\alpha}$.
Therefore, $J^{+-}\Gamma$ approaches constant, 
giving a finite minimum conductivity in the limit of $\alpha\to0$.

A finite conductivity at absolutely no scattering $(\alpha=0)$ 
looks counterintuitive,
but here we should note that the result is based on 
the implicit assumption that
the transport is diffusive,
i.e. the system size is much greater than the mean free path.
If we take a limit  $\alpha\to 0$  in a fixed-sized system, 
the mean free path exceeds the system size at some point and
then the diffusive transport switches to the ballistic transport,
to which the present conductivity formula does not apply.

\begin{figure}
\begin{center}
\leavevmode\includegraphics[width=0.9\hsize]{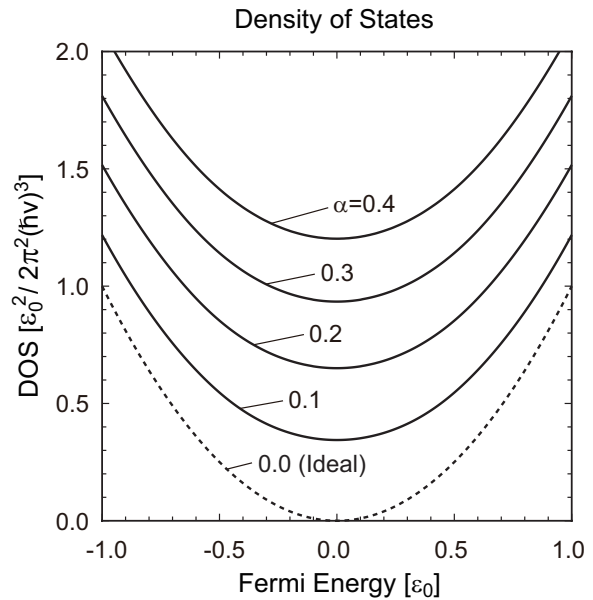}
\end{center}
\caption{Density of states calculated by the SCBA,
as a function of the Fermi energy
at several values of $\alpha$.
}
\label{dos_fig}
\end{figure}

\begin{figure}
\begin{center}
\leavevmode\includegraphics[width=0.9\hsize]{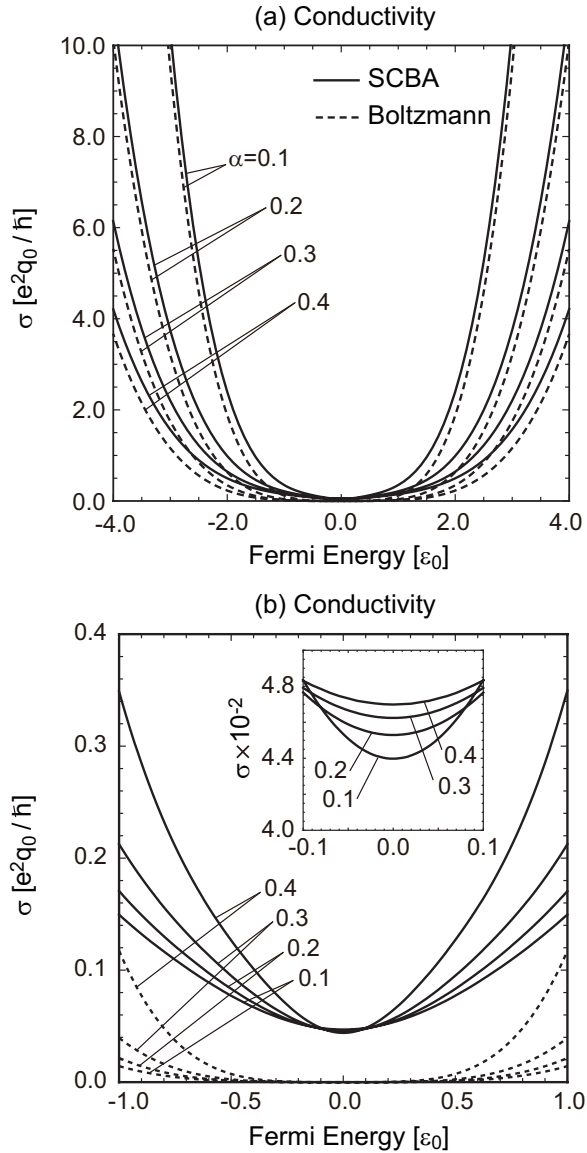}
\end{center}
\caption{Conductivity as a function of
the Fermi energy in different plot ranges. 
In each panel, the solid lines represent the SCBA  
and the dashed lines represent the Boltzmann theory.
The inset in (b) shows the detailed plot for the SCBA conductivity
around the Weyl point.
}
\label{cond_fig}
\end{figure}

\begin{figure}
\begin{center}
\leavevmode\includegraphics[width=0.9\hsize]{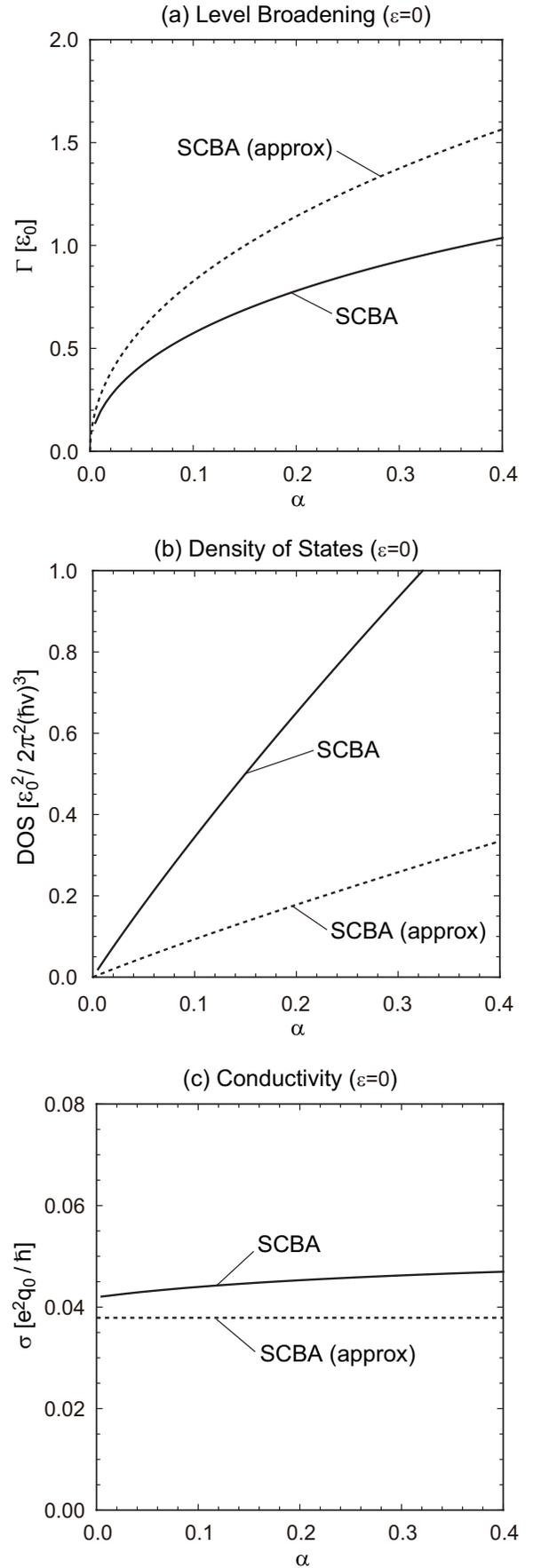}
\end{center}
\caption{Density of states and the conductivity
as a function of $\alpha$.
The solid line represent the numerical result
and the dashed line represent the approximate
analytical expression (see text).
}
\label{cond_dos_fig}
\end{figure}

\section{Numerical Results}
\label{sec_num}


We solve the SCBA equations Eq.\ (\ref{eq_self_x}),
Eq.\ (\ref{eq_self_y}), and (\ref{eq_self_J})
by numerical iteration and calculate the density of states 
and the conductivity.
Figure \ref{dos_fig} shows the density of states
as a function of the Fermi energy
at several values of $\alpha$.
The density of states is enhanced in all energy region
linearly to $\alpha$, and this is consistent with
the behavior in the analytical expression at $\e=0$
in the previous section [Eqs.\ (\ref{dos_at_zero}) and (\ref{q_g_approx})].
Fig.\ \ref{cond_dos_fig}(a) shows the level broadening
$\Gamma = {\rm Im}[X(k=0,\varepsilon=0)]$
as a function of $\alpha$, where the solid line shows
the numerical result and the dashed line shows
the approximate solution Eq.\ (\ref{q_g_approx}).
Fig.\ \ref{cond_dos_fig}(b) is a similar plot for the density of states
$D(\varepsilon=0)$ as a function of $\alpha$,
where the solid line represent the numerical result and
the dashed line represent the approximate analytical expression
Eq.\ (\ref{dos_at_zero}).
We see that in the both plots
the analytical expression well reproduces
the qualitative behavior of the numerical result,
i.e., $\Gamma\propto\sqrt{\alpha}$ and $D(0)\propto \alpha$.


Figs. \ref{cond_fig}(a) and (b) shows the conductivity
as a function of the Fermi energy
for several values of $\alpha$.
In Fig.\ \ref{cond_fig}(a), we see that
the SCBA result mostly agrees with
the Boltzmann theory away from $\e=0$,
where the conductivity is proportional to $\e^4$
and increases with the decrease of $\alpha$
as expected from Eq.\ (\ref{Boltzmann}).
Fig.\ \ref{cond_fig}(b) shows the detailed plot
around the Weyl point.
Now we see a considerable disagreement between the two results,
where the Boltzmann conductivity vanishes at the Weyl point,
although the SCBA conductivity has a finite value.
The Boltzmann theory is valid
when the Fermi energy is much greater
than the level broadening $\Gamma$, 
so that the energy region where the Boltzmann theory fails
becomes wider with the increase of $\alpha$.
We actually see this behavior in Figs.\ \ref{cond_fig}(a).

Fig.\ \ref{cond_dos_fig}(c) shows the zero-energy conductivity $\sigma(0)$
as a function of $\alpha$,
where the solid line indicates the numerical result and the dashed line
the analytical expression Eq.\ (\ref{cond_approx2}).
The numerical curve is nearly constant depending on $\alpha$
 only weakly.
In the limit of $\alpha\to 0$, it actually approaches a finite value, 
and the magnitude agrees qualitatively well 
with the analytic estimation of Eq.\ (\ref{cond_approx2}).



\section{Discussion}
\label{sec_dis}

\subsection{Validity of SCBA at the Weyl point}

\begin{figure}
\begin{center}
\leavevmode\includegraphics[width=0.9\hsize]{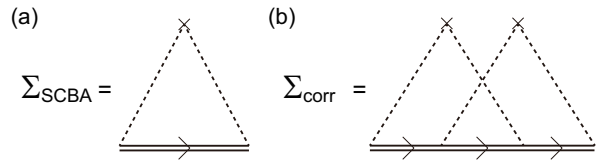}
\end{center}
\caption{The diagrammatic representations
of (a) the self-energy for the SCBA and
(b) the leading correction term for the SCBA. 
}
\label{diagram_self_fig}
\end{figure}

Since the SCBA only partially takes the self-enegy diagrams
in the perturbational expansion, it is generally suppose to be valid
when the scattering strength is relatively weak.
Fig.\ \ref{diagram_self_fig} (a) expresses
SCBA self-energy $\Sigma_{\rm SCBA}$, 
and (b) shows the leading correction term $\Sigma_{\rm corr}$ 
which was neglected in the SCBA.
The SCBA is qualitatively correct when
$\Sigma_{\rm corr}$ is much smaller than $\Sigma_{\rm SCBA}$.
In the conventional disordered metal, we have 
$\Sigma_{\rm corr}/\Sigma_{\rm SCBA} = \mathcal{O}(1/k_{\rm F}l)$ with the Fermi wave vector $k_{\rm F}$
and the mean free path $l$.

It is nontrivial if the SCBA is valid at the Weyl point
where $k_F$ becomes zero.\cite{ostrovsky2006electron,PhysRevLett.113.026602}
In the presence of the disorder potential, 
$k_F$ does not actually vanish but it is effectively replaced with
$\sim \Gamma/(\hbar v)$ due to the finite level broadening $\Gamma$.
Meanwhile the mean free path $l$ is given by $v \tau$ where $v$ is the 
constant band velocity and $\tau = \hbar/\Gamma$ is the scattering time.
Then we end up with $k_F l = \mathcal{O}(1)$,
which means the correction term is not actually negligible.

In a recent theoretical study \cite{PhysRevLett.113.026602},
the conductivity in the single-node 3D Weyl electron 
is numerically calculated in the presence of the Gaussian impurities
using the Landuer formulation.
The behaviors of the Weyl-point self-energy and conductivity 
are found to be consistent 
with the corresponding SCBA calculation \cite{ominato2014quantum},
while there is a quantitative discrepancy by a factor.
Ref.\ \onlinecite{PhysRevLett.113.026602} also estimated the leading correction term 
$\Sigma_{\rm corr}$ in the numerical calculation and it
was found to be smaller than $\Sigma_{\rm SCBA}$ 
but not negligibly small. This is actually responsible for the quantitative 
descrepancy in the SCBA.

In the following, we consider the extended SCBA approximation
including the leading correction term $\Sigma_{\rm corr}$ 
for the screened Coulomb impurity case, 
and show that the additonal term does not change the qualitative behevior
of the total self-energy.
The extended self-consistent equation including the diagrams of
Fig.\ \ref{diagram_self_fig} (a) and (b) is written as
\begin{align}
\hat \Sigma&(\bm{k},\e)=\int\frac{{\rm d}\bm{k}^\p}{(2\pi)^3}n_{\rm i}
                                   |u(\bm{k}-\bm{k}^\p)|^2\hat G(\bm{k}^\p,\e) \notag \\
                &+\int\frac{{\rm d}\bm{k}^\p}{(2\pi)^3}\int\frac{{\rm d}\bm{k}^{\p\p}}{(2\pi)^3}n_{\rm i}^2
                   |u(\bm{k}-\bm{k}^\p)|^2|u(\bm{k}^\p-\bm{k}^{\p\p})|^2 \notag \\
                &\hspace{1cm}\times\hat G(\bm{k}^\p,\e)\hat G(\bm{k}^{\p\p},\e)
                                             \hat G(\bm{k}-\bm{k}^\p+\bm{k}^{\p\p},\e).
\label{eq_self_discus}
\end{align}
We consider the Weyl point $\varepsilon=0$
and assume $\hat \Sigma=-i\Gamma$ and $q_{\rm s}\ll\Gamma/(\hbar v)$ as done in Sec.\ \ref{sec_app}.
As the $u(\bm{k})$ term is relevant only when $k <\sim q_s$,
we can replace the Green's function $\hat{G}(\bm{k})$ with $1/(i\Gamma)$
under the present assumption $k <\sim q_{\rm s}\ll\Gamma/(\hbar v)$.
Then $k$-integral simply gives $I$ of Eq.\ (\ref{eq_I_def}),
and the self-consistent equation\ (\ref{eq_self_discus}) is reduced to
\begin{align}
\Gamma=\Gamma\left[
                         \left(\frac{\Gamma_0}{\Gamma}\right)^2
                         +\left(\frac{\Gamma_0}{\Gamma}\right)^4
                         \right],
                         \label{eq_self_corr}
\end{align}
where $\Gamma_0$ is defined in Eq.\ (\ref{gamma0}).
By solving this, we find a non-trivial solution
\begin{align}
\Gamma=\sqrt{\frac{1+\sqrt{5}}{2}}\Gamma_0.
\label{gamma_correction}
\end{align}
The ratio of the second term to the first term
in Eq.\ (\ref{eq_self_corr}) then gives
\begin{align}
\frac{\Sigma_{\rm corr}}{\Sigma_{\rm SCBA}} = 
\left(\frac{\Gamma_0}{\Gamma}\right)^2
=\frac{-1+\sqrt{5}}{2} \approx 0.618\cdots,
\label{eq_corr_scba}
\end{align}
i.e., $\Sigma_{\rm corr}$ is smaller than $\Sigma_{\rm SCBA}$ 
while not negligibly small. 
In fact,
Eq.\ (\ref{eq_corr_scba}) is close to the value numerical estimated for
the Gaussian impurity case in Ref. \onlinecite{PhysRevLett.113.026602}.

In the usual SCBA approach without $\Sigma_{\rm corr}$
in the previous section, 
we only take the first term in the bracket of Eq.\ (\ref{eq_self_corr})
and obtain $\Gamma=\Gamma_0$.
Comparing to Eq.\ (\ref{gamma_correction}), 
we see that the correction term attaches a numerical factor
in front of the SCBA self-energy. Therefore, we expect that 
adding the correction terms does not change the qualitative behavior
of the total self-energy.


\subsection{Critical behavior in a general impurity potential under the screening effect}

In our previous work, we studied the quantum transport
in 3D Weyl electron in presence of Gaussian impurities,
i.e., impurity potential $U(\textbf{r})$
expressed by a Gaussian $U_0 \exp(-r^2/r_0^2)$.
\cite{ominato2014quantum}
There it was found that the density of states and the conductivity
at the Weyl point completely vanish below a certain critical disorder strength, 
and abruptly rise above it.\cite{ominato2014quantum}
On the other hand, 
we also showed that such a critical behavior is never observed
in the bare (i.e., unscreened) Coulomb potential,
and the absence of the critical point is attributed to 
the divergence of $u(q)$ in the limit of $q\to 0$.\cite{ominato2014quantum}

Unlike the bare Coulomb potential,
the screened Coulomb potential studied in this paper
does not diverge in $q\to 0$ due to the finite screening length, and then
we naively expect the critical behavior takes place in a similar way to 
Gaussian impurities.
Contrary to such an expectation, the detailed calculation in the above section
showed no critical behaviors in the screened Coulomb impurity.
To resolve this apparent discrepancy,
we argue in the following about the criteria for the 
critical behavior 
in general impurity potential with the screening effect.

We consider the isotropic impurity potential
$U(r)$ (and its Fourier tranform $u(k)$),
and assume an approximate solution 
for the self-consistent equation,
\begin{align}
X(k,0)&=i\Gamma, \\
Y(k,0)&=\hbar v k.
\end{align}
Then Eq.\ (29) at $k=0$ is written as
\begin{align}
\Gamma=\frac{n_{\rm i}}{2\pi^2}\int^\infty_0{k^\p}^2{\rm d}k^\p
u(k^\p)^2
\frac{\Gamma}{{\Gamma^2+(\hbar v k^\p)^2}}.
\label{eq_self_gamma_org}
\end{align}
Obviously, Eq.\ (\ref{eq_self_gamma_org}) has a trivial solution $\Gamma=0$,
and another solution is obtained from
\begin{align}
1=\frac{n_{\rm i}}{2\pi^2}\int^\infty_0{k^\p}^2{\rm d}k^\p
u(k^\p)^2\frac{1}{{\Gamma^2+(\hbar v k^\p)^2}}.
\label{eq_self_gamma}
\end{align}
When the right-hand side of Eq.\ (\ref{eq_self_gamma}) 
is viewed as a function of $\Gamma$, 
it takes the maximum value at $\Gamma= 0$, which is written as,
\begin{align}
	I=\frac{n_{\rm i}}{2\pi^2\hbar^2 v^2}
	\int^\infty_0{\rm d}k^\p
	u(k^\p)^2.
	\label{eq_I_def}
\end{align}
When $I$ is smaller than 1, Eq.\ (\ref{eq_self_gamma}) 
cannot be satisfied by any $\Gamma$, and then
$\Gamma= 0$ is the only solution of
Eq.\ (\ref{eq_self_gamma_org}).
In the case of the Gaussian potential 
$u(k) =u_0\exp(-k^2/k_0^2)$, for example,
the integral $I$ becomes a finite value
proportional to $n_{\rm i}u_0^2$,
and $\Gamma$ (and thus the density of states) vanishes
when $n_{\rm i}u_0^2$ is lower than a certain critical value.
\cite{ominato2014quantum}

For the screened Coulomb potential, 
i.e., 
$u(k) =(4\pi e^2/\kappa) / ({k^\p}^2+q_{\rm s}^2)$,
we have
\begin{align}
I
= 2\pi\alpha^2\left(\frac{q_0}{q_{\rm s}}\right)^3,
\label{eq_I_coulomb}
\end{align}
and the condition for having only a trivial solution $\Gamma=0$ is
\begin{align}
1\ge2\pi\alpha^2\left(\frac{q_0}{q_{\rm s}}\right)^3.
\label{eq_cond_coulomb}
\end{align}
If we treat $q_{\rm s}$ as a constant,
Eq.\ (\ref{eq_cond_coulomb})
is satisfied when $\alpha$ is sufficiently small.
However, $\alpha$ and $q_{\rm s}$ are not actually independent
in the self-consistent calculation, as we argued in Sec.\ \ref{sec_app}.
Using the self-consistent solution Eqs.\ (\ref{eq_TF2}) and (\ref{q_g_approx}),
Eq.\ (\ref{eq_cond_coulomb}) is rewritten as
\begin{align}
1 \geq 
\left(
1+\frac{1}{\sqrt{f\alpha}}
\right)^2,
\end{align}
which cannot be true.  
In a screened Coulomb scatterers, therefore,
we always have a nonzero solution for $\Gamma$
and there is no critical disorder scattering strength.

On the other hand, we can show that the critical disorder strength does 
exist in Gaussian scatterers even when including the screening effect,
which was neglected in the previous work.\cite{ominato2014quantum}
The screened Gaussian potential is written as\cite{ominato2014quantum}
\begin{align}
u(k)=\frac{u_0\exp(-k^2/k_0^2)}{1+q_{\rm s}^2/k^2},
\end{align}
giving
\begin{align}
I=\frac{n_{\rm i}u_0^2}{2\pi^2\hbar^2 v^2}
&\int^\infty_0{\rm d}k^\p   
\left(\frac{\exp(-{k^\p}^2/k_0^2)}{1+q_{\rm s}^2/{k^\p}^2}\right)^2.
\end{align}
The inverse screening length $q_{\rm s}$ is to be self-consistently determined 
by Eq.\ (\ref{eq_TF}).
Unlike the Coulomb impurity [Eq.\ (\ref{eq_I_coulomb})], 
the intergral $I$ never diverges in any value of $q_{\rm s}$
and it has an upper bound $I_{\rm max}$ at $q_{\rm s} =0$. 
In a sufficiently small $n_{\rm i}u_0^2$
such that $I_{\rm max}<1$, therefore,
we have only a trivial solution $\Gamma=0$ regardless of $q_{\rm s}$,
while this is a sufficient but not necessary condition.

Following the above discussion, 
we see that whether a critical disorder strength exists
depends on the specific form of the impurity potential,
even when the screening effect is included.
We can examine the existence of the critical disorder strength 
for any type of impurity scatterers
in a similar way,
by estimating the maximum value of the intergral $I$ in Eq.\ (\ref{eq_I_def})
as a function of $q_{\rm s}$.

\section{Conclusion}
\label{sec_conc}

We have studied the electronic transport in
three-dimensional Weyl electron system
with the charged Coulomb impurities
using the self-consistent Born approximation.
The scattering strength is characterized
by the effective fine structure constant $\alpha$
which is determined by the Fermi velocity and
the dielectric constant.
The density of states is enhanced in all energy region
and at a fixed energy, it increases linearly with the increase of $\alpha$.
On the other hand the conductivity at the Weyl point is almost independent of
$\alpha$, and even survive in the limit of $\alpha\to0$.
The magneitude of the Weyl-point conductivity only depends on the impurity 
density $n_{\rm i}$, and scales in proportion to $n_{\rm i}^{1/3}$.
In the energy region away from the Weyl point,
the SCBA conductivity agrees well with the Boltzmann conductivity.
The behavior in Coulomb impurities 
is significantly different from the Gaussian impurities,
where the Weyl point conductivity almost completely vanishes
below a finite critical disorder strength.
We showed that the existence of the critical disorder strength
can be tested by an analytic criteria
for the impurity potential $U(r)$.

\section*{ACKNOWLEDGMENTS}


\appendix*

\section{Self-consistent Born approximation}

Here we present the derivation of the self-consistent equations
and the formula for the conductivity.
Using the definition of $X(k,\e)$ and $Y(k,\e)$,
Eqs.\ (\ref{def_Gf}) and (\ref{eq_self}) are written as
\begin{align}
\hat G(\bm{k},\e)= \frac{1}{X(k,\e)-Y(k,\e)(\bm{\sigma}\cdot\bm{n})},
\end{align}
and 
\begin{align}
\hat \Sigma(\bm{k},\e)=\int\frac{{\rm d}\bm{k}^\p}{(2\pi)^3}n_{\rm i}|u(\bm{k}-\bm{k}^\p)|^2
                           \frac{X^\p+Y^\p(\bm{\sigma}\cdot\bm{n}^\p)}{{{X^\p}^2-{Y^\p}^2}} \label{self1}
\end{align}
where $X^\p=X(k^\p,\e)$, $Y^\p=Y(k^\p,\e)$, and $\bm{n}^\p=\bm{k}^\p/k^\p$.

Now, we divide $\bm{n}^\p$ as
\begin{align}
\bm{n}^\p=\bm{n}^\p_\parallel+\bm{n}^\p_\perp.
\end{align}
where $\bm{n}^\p_\parallel=(\bm{n}\cdot\bm{n}^\p)\bm{n}$ 
is the component of parallel to $\bm{n}$,
and $\bm{n}^\p_\perp$ is the perpendicular part.
Then Eq.\ (\ref{self1}) becomes
\begin{align}
\hat \Sigma(\bm{k},\e)=&\int\frac{{\rm d}\bm{k}^\p}{(2\pi)^3}n_{\rm i}|u(\bm{k}-\bm{k}^\p)|^2
                                      \frac{X^\p}{{{X^\p}^2-{Y^\p}^2}} \notag \\
                                 +&\int\frac{{\rm d}\bm{k}^\p}{(2\pi)^3}n_{\rm i}|u(\bm{k}-\bm{k}^\p)|^2
                                      \frac{Y^\p}{{{X^\p}^2-{Y^\p}^2}}(\bm{\s}\cdot\bm{n}^\p_\parallel) \notag \\
                                 +&\int\frac{{\rm d}\bm{k}^\p}{(2\pi)^3}n_{\rm i}|u(\bm{k}-\bm{k}^\p)|^2
                                      \frac{Y^\p}{{{X^\p}^2-{Y^\p}^2}}(\bm{\s}\cdot\bm{n}^\p_\perp).
\end{align}
The third term vanishes after the integration over the $\bm{k}^\p$ direction,
giving
\begin{align}
\hat \Sigma(\bm{k},\e)=&\int_0^\infty\frac{{k^\p}^2{\rm d}k^\p}{(2\pi)^3}n_{\rm i}V_0^2(k,k^\p)
                                      \frac{X^\p}{{{X^\p}^2-{Y^\p}^2}} \notag \\
                                 +&(\bm{\sigma}\cdot\bm{n})
                                     \int_0^\infty\frac{{k^\p}^2{\rm d}k^\p}{(2\pi)^3}n_{\rm i}V_1^2(k,k^\p)
                                      \frac{Y^\p}{{{X^\p}^2-{Y^\p}^2}}.
\label{self2}
\end{align}
The above equation immediately gives
the self-consistent equation Eq.\ (\ref{eq_self_x}) and (\ref{eq_self_y}).

The Kubo formula for the conductivity is given by
\begin{align}
\sigma(\e)&=-\frac{\hbar e^2v^2}{4\pi}\sum_{s,s^\p=\pm1}s
 s^\p\int\frac{{\rm d}\bm{k}^\p}{(2\pi)^3}  
{\rm Tr}\biggl[
\s_x\hat G(\bm{k}^\p,\e+is0) \notag \\
&{~~~}\times\hat J_x(\bm{k}^\p,\e+is0,\e+is^\p0)\hat
 G(\bm{k}^\p,\e+is^\p0)
\biggr],
\end{align}
where $\hat J_x$ is current vertex-part satisfying the Bethe-Salpeter equation
\begin{align}
\hat J_x(\bm{k},\e,\e^\p)=\sigma_x+&\int\frac{{\rm d}\bm{k}^\p}{(2\pi)^3}
                                                   n_{\rm i}|u(\bm{k}-\bm{k}^\p)|^2\hat G(\bm{k}^\p,\e) \notag \\
&\times\hat J_x(\bm{k}^\p,\e,\e^\p)\hat G(\bm{k}^\p,\e^\p).
\label{GJG}
\end{align}
The vertex part $\hat J$ is written as
\begin{align}
\hat J_x(\bm{k},\e,\e^\p)=\s_x J_0(k,\e,\e^\p)+(\bm{\s}\cdot\bm{n})\s_x(\bm{\s}\cdot\bm{n}) J_1(k,\e,\e^\p) \notag \\
                           +(\bm{\s}\cdot\bm{n})\s_x J_2(k,\e,\e^\p)+\s_x(\bm{\s}\cdot\bm{n})J_3(k,\e,\e^\p).
\label{J}
\end{align}
To calculate Eq.\ (\ref{GJG}), we consider an integral
\begin{align}
I(\bm{k})=\int\frac{{\rm d}\bm{k}^\p}{(2\pi)^3}|u(\bm{k}-\bm{k}^\p)|^2F(k^\p)
                   (\bm{\s}\cdot\bm{n}^\p)\s_x(\bm{\s}\cdot\bm{n}^\p),
\end{align}
where $F(k)$ is an arbitrary function.
After some algebra, we obtain
\begin{align}
I(\bm{k})=&\s_x\int\frac{{k^\p}^2{\rm d}k^\p}{(2\pi)^3}F(k^\p) 
\left(-\frac{1}{2}V_0^2(k,k^\p)+\frac{1}{2}V_2^2(k,k^\p)\right) \notag \\
             &+(\bm{\s}\cdot\bm{n})\s_x(\bm{\s}\cdot\bm{n})\int\frac{{k^\p}^2{\rm d}k^\p}{(2\pi)^3}F(k^\p) \notag \\
& \qquad\qquad
\times\left(-\frac{1}{2}V_0^2(k,k^\p)+\frac{3}{2}V_2^2(k,k^\p)\right).
\end{align}
In a similar way as for the self-energy, we have
\begin{align}
 &\int\frac{{\rm
 d}\bm{k}^\p}{(2\pi)^3}|u(\bm{k}-\bm{k}^\p)|^2F(k^\p)(\bm{\s}\cdot\bm{n}^\p)\s_x
 \notag \\
& \qquad =(\bm{\s}\cdot\bm{n})\s_x\int\frac{{k^\p}^2{\rm d}k^\p}{(2\pi)^3}F(k^\p)V_1^2(k,k^\p), \notag \\
 &\int\frac{{\rm d}\bm{k}^\p}{(2\pi)^3}|u(\bm{k}-\bm{k}^\p)|^2F(k^\p)\s_x(\bm{\s}\cdot\bm{n}^\p) \notag \\
& \qquad = \s_x(\bm{\s}\cdot\bm{n})\int\frac{{k^\p}^2{\rm d}k^\p}{(2\pi)^3}F(k^\p)V_1^2(k,k^\p).
\end{align}
Using the above equations, we obtain the Bethe-Salpeter equation Eq.\ (\ref{eq_self_J})
and Eq.\ (\ref{eq_cond}).

\bibliography{3d_weyl_dc_charged}

\end{document}